\documentclass{aa}

\usepackage{graphicx}
\usepackage{natbib}
\usepackage{longtable}
\usepackage{txfonts}
\usepackage{color}
\usepackage[T1]{fontenc}
\usepackage{ulem}   

\usepackage{gensymb}

\usepackage[flushleft]{threeparttable}

\def\lum{erg s$^{-1}$}

\def\xte{{\it RXTE}}

\def\gro{GRO~J1744$-$28}

\newcommand{\subt}[1]{_{\text{#1}}}

\newcommand {\be}{\begin {equation}}
\newcommand {\ee}{\end {equation}}
\newcommand {\beq}{\begin {eqnarray}}
\newcommand {\eeq}{\end {eqnarray}}

\newcommand{\powten}[1]{\times 10^{#1}}

\begin{document}

   \title{Evidence for the radiation-pressure dominated accretion disk in bursting pulsar GRO~J1744$-$28 using timing analysis}

   \author{Juhani M\"onkk\"onen \inst{1}
   		  \and Sergey S. Tsygankov \inst{1,2}
          \and Alexander A. Mushtukov \inst{3,2,4}
          \and Victor Doroshenko \inst{5}
          \and Valery F. Suleimanov \inst{5,6,2}
          \and Juri Poutanen \inst{1,2,7}
          }

   \institute{Department of Physics and Astronomy, FI-20014 University of Turku, Finland          
       \and
       Space Research Institute of the Russian Academy of Sciences, Profsoyuznaya Str. 84/32, Moscow 117997, Russia    
       \and 
       Leiden Observatory, Leiden University, NL-2300RA Leiden, The Netherlands
       \and
       Anton Pannekoek Institute, University of Amsterdam, Science Park 904, 1098 XH Amsterdam, The Netherlands
       \and
       Institut f\"ur Astronomie und Astrophysik, Universit\"at T\"ubingen, Sand 1, D-72076 T\"ubingen, Germany
       \and
       Kazan (Volga region) Federal University, Kremlevskaya str. 18, 420008 Kazan, Russia
          \and Nordita, KTH Royal Institute of Technology and Stockholm University, Roslagstullsbacken 23, SE-10691 Stockholm, Sweden
          }
   \titlerunning{Radiation-pressure dominated accretion disk in \gro}
   \authorrunning{ M\"onkk\"onen et al. }
   \date{  }


  \abstract
      { The X-ray pulsar \gro\ is a unique source which shows both pulsations and type-II X-ray bursts, allowing studies of the interaction of the accretion disk with the magnetosphere at huge mass accretion rates exceeding $10^{19}$~g s$^{-1}$ during its super-Eddington outbursts. The magnetic field strength in the source, $B\approx 5\times 10^{11}$~G, is known from the cyclotron absorption feature discovered in the energy spectrum around 4.5 keV. Here, we explore the flux variability of the source in context of interaction of its magnetosphere with the radiation-pressure dominated accretion disk. Particularly, we present the results of the analysis of noise power density spectra (PDS) using the observations of the source in 1996--1997 by the \textit{Rossi X-ray Timing Explorer} ({\it RXTE}). Accreting compact objects commonly exhibit a broken power-law shape of the PDS with a break corresponding to the Keplerian orbital frequency of matter at the innermost disk radius. The observed frequency of the break can thus be used to estimate the size of the magnetosphere. We found, however, that the observed PDS of \gro\ differs dramatically from the canonical shape. Furthermore, the observed break frequency appears to be significantly higher than what is expected based on the magnetic field estimated from the cyclotron line energy. We argue that these observational facts can be attributed to the existence of the radiation-pressure dominated region in the accretion disk at luminosities above $\sim$2$\powten{37}$~\lum. We discuss a qualitative model for the PDS formation in such disks, and show that its predictions are consistent with our observational findings. The presence of the radiation-pressure dominated region can also explain the observed weak luminosity-dependence of the inner radius, and we argue that the small inner radius can be explained by a quadrupole component dominating the magnetic field of the neutron star.}  

   \keywords{accretion, accretion disks
             -- magnetic fields
             -- stars: individual: GRO J1744$-$28
             -- X-rays: binaries
             -- stars: neutron
               }

   \maketitle

%


\section{Introduction}

\gro\ is a transient X-ray pulsar (XRP) discovered in December 1995 with the Burst and Transient Source Experiment (BATSE) on \textit{Compton Gamma-Ray Observatory} \citep{Fishman1995}. It showed frequent, spectrally hard bursts and the source was suspected to be an accreting neutron star (NS) with type II bursts, that is, bursts originating from accretion instabilities \citep{Kouveliotou1996}. Persistent pulsations with a period of 467\,ms and subsequent spin-up confirmed that the source is indeed a magnetised NS accreting from a low-mass companion star via a disk. The binary orbital period was measured to be 11.8 days \citep{Finger1996}. Since the source was the first to show both bursts and pulsations, it was given the nickname ``Bursting Pulsar''. The attempts of observing the companion have been inconclusive \citep{Cole1997,Gosling2007}, therefore the distance to the source is rather uncertain. The hydrogen column density was, however, measured from spectra to be around $5\powten{22}$~cm$^{-2}$, which is consistent with typical values for sources located in the Galactic center, and thus the distance to the source is usually assumed to be around 8~kpc \citep{Dotani1996, Nishiuchi1999}.  For that distance  the apparent luminosities at the peaks of major outbursts exceed the Eddington luminosity for a NS. 
 
Recently, \citet{DAi2015} found a cyclotron absorption line at $4.7$~keV in the spectrum obtained with EPIC/pn  instrument on board of {\it XMM-Newton}, indicating the surface magnetic field strength of $B= (5.27 \pm 0.06) \powten{11}$~G. The same feature was independently discovered by \citet{Doroshenko2015} in the older {\it BeppoSAX} data. Thus \gro\ has a magnetic field of intermediate strength when compared to other accreting magnetized NSs \citep[see a review by][]{Staubert2019}. This, together with the observed super-Eddington outbursts, implies that the magnetospheric radius and thus the inner radius of the accretion disk should be rather small, around $5\times10^7$~cm, which makes the Bursting Pulsar a unique object to study, particularly, in the context of super-Eddington accretion onto a NS as a mechanism to power ultraluminous X-ray sources. 

Analysis of flux variability has proven to be a valuable tool for the study of accretion physics. The observed variability in the X-ray flux and therefore in the mass accretion rate is linked to the viscous processes in the accretion disk, which may be described by the perturbation propagation model \citep{Lyubarskii1997,Kotov2001, Churazov2001, Arevalo2006, Ingram2013, Mushtukov2018,2019MNRAS.tmp..913M}.     In the case of an extremely high mass accretion rate, the flux variability can be modified by an optically thick envelope formed by the accretion flow covering the magnetosphere of a neutron star \citep{2017MNRAS.467.1202M,2019MNRAS.484..687M}.
The resulting power spectra of X-ray flux variability have a power-law shape, or a broken power-law shape if the accretion disk is truncated at some point. For instance, XRPs have a well-defined inner edge of the accretion disk coinciding with the magnetospheric radius. Due to the magnetospheric truncation of the disk, the power spectrum continuum breaks to a steeper slope, which has an index around $-2$. The break frequency is related to the maximal frequency generated in the disk. It was noted already by \citet{Hoshino1993} that in accreting XRPs the high-frequency break typically coincides with the frequency of regular pulsations. \citet{Revnivtsev2009} examined PDS of several XRPs which are accreting in equilibrium and found that all of the sources have a break in power law close to the source spin frequency. In these sources, which are believed to be close to corotation, the disk is expected to be truncated at the corotation radius, where the orbital frequency of the disk matter coincides with the pulsar spin frequency \citep[for a more detailed discussion about the torque equilibrium and the inner radius see][]{Aly1990, Lovelace1995, Li1999}. Therefore, the break frequency can be taken to be closely related to the Keplerian frequency of the matter at the innermost radius. 

Moreover, \citet{Revnivtsev2009} demonstrated that in transient XRPs characterized by large variations in the mass accretion rate, the break frequency as a function of the X-ray flux follows the classical dependence of the Alfv\'en radius on the accretion rate. Based on this connection between the break frequency and the Keplerian frequency of the magnetospheric radius, a novel method for determination of the magnetic field in such sources was proposed and successfully applied to accreting NSs and white dwarfs \citep[see, e.g.,][]{Revnivtsev2010,2012MNRAS.421.2407T,2014A&A...561A..96D,2019MNRAS.482.3622S}. 

In this paper, we analyze the data obtained during the two major outbursts of \gro\ in 1996 and 1997 in order to explore the PDS of accreting magnetized NS with a comparatively weak magnetic field, and, in the light of the obtained results, attempt
to generalize the result of \citet{Revnivtsev2009} to the case of a radiation-pressure dominated accretion disk expected in this unique source. The description of the data, their reduction and the analysis are given in Sect.~\ref{sec:data}. The results of the extensive timing analysis are described in Sect.~\ref{sec:results}.
In Sect.~\ref{sec:discuss} we discuss the implications of our results, and we conclude in Sect.~\ref{sec:conclusions}.

\section{Observations, data reduction and analysis}
\label{sec:data}

\begin{figure}
\centering
\includegraphics[width=\columnwidth]{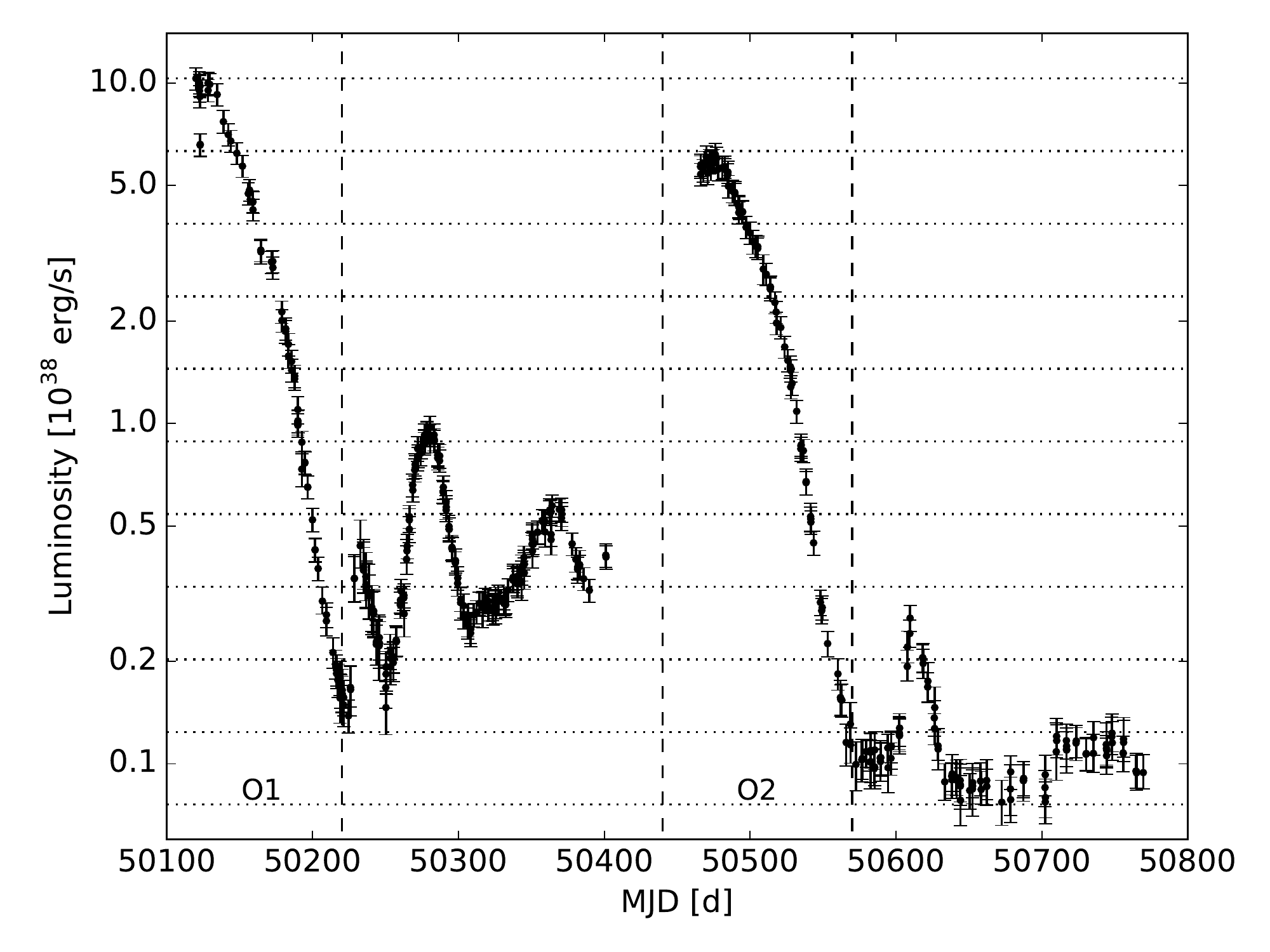} 
\caption{Long-term X-ray light curve of \gro\ with luminosities calculated in the 2.5--45 keV energy band. The horizontal dotted lines indicate the luminosity bins, in which the PDS were averaged together. The vertical dashed lines divide the light curve into separate time intervals; O1 and O2 mark the 1st and 2nd outbursts analyzed here.}
\label{fig:lc_entire}
\end{figure}

From January 1996 to November 1997, during the first two bright outbursts, \gro\ was observed by the {\it Rossi X-Ray Timing Explorer} (\xte) more or less regularly, covering the source in a wide range of luminosities as seen in Figure \ref{fig:lc_entire}. Proportional Counter Array (PCA) data were chosen for our research because of the large effective area of the instrument and its high time resolution, which allow the construction of power spectra over a wide frequency range with good statistics. The instrument has a large field of view with a radius of $1^\circ$, which comes with the downside that other X-ray sources can contaminate the observations, especially when the source flux is low. This may happen with \gro\ because of its location in the direction of the highly populated center of our Galaxy. However, extremely high luminosity of \gro\ during the outbursts still enables us to study its properties with certainty.
 
In the analysis, we focus on the PDS of the two major outbursts. We exclude the ``mini-outbursts'' occurring after the major outbursts from our analysis because changes in several observational properties indicate that the conditions in the disk-magnetosphere interaction are different: the luminosities are lower and the type II bursting behavior becomes different \citep{Stark1998}. Recently, a thorough classification of the type II bursts and an analysis of the bursting behavior was presented by \citet{Court2018a}. In addition to those changes, the pulse profile changes from the highly sinusoidal seen in outbursts \citep{Giles1996} to a much more irregular in the mini-outbursts, which indicates a change in the accretion geometry. Because of a wide field of view, separating the effect of other X-ray sources is also challenging at low luminosities.

\subsection{Timing analysis}
\label{sec:timing}

Extraction of the light curves and spectra from PCA data files was done with tools from {\sc heasoft} software package\footnote{https://heasarc.nasa.gov/lheasoft/} v. 6.19 and following the standard procedures of {\it RXTE} Cook Book guide.\footnote{https://heasarc.gsfc.nasa.gov/docs/xte/recipes/cook\_book.html} For our analysis, the data configuration with the best time resolution was chosen from each observation. During the outburst decay periods the configuration was almost entirely in the single-bit mode in science array format, separately observed in two energy bands (approximately 1$-$10 keV and 10$-$100 keV) and with a time resolution of $244\ \mu\text{s}$. The single-bit mode combines data from all the Proportional Counter Units (PCUs). The other chosen files were mostly in the science event format with a resolution of $122\ \mu\text{s}$, collected over the entire \xte/PCA energy range, and with some of the PCUs, except PCU2, turned off from time to time. Thus, for the science event configuration only PCU2 data was used for consistency.  
 
\begin{figure}
\centering
\includegraphics[width=\columnwidth]{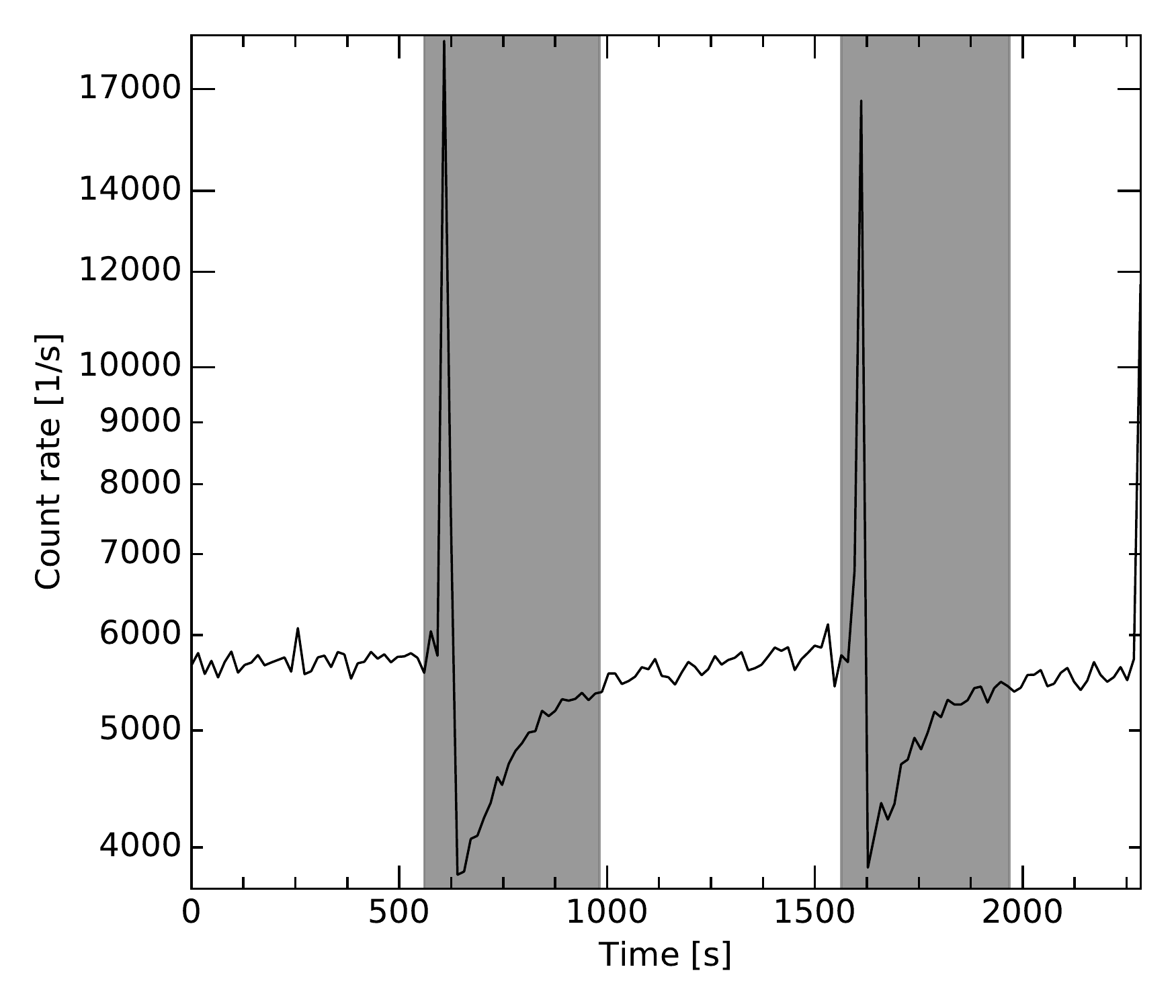} 
\caption{X-ray light curve starting at MJD 50138.81 during the peak of the first outburst showing type II bursts. Only the periods with nearly constant flux level were included in the analysis (white area),  while the burst and depression periods after them were excluded (dark area). 
}\label{fig:lc_example}
\end{figure}

The type II bursts can be seen in the example light curve from the peak of the first outburst shown in Figure \ref{fig:lc_example}. The light curve can be divided into three phases corresponding to a nearly constant flux, a short-duration burst and a dip or depression with a recovery back to the constant level. The differences between the phases have been analyzed in the literature \citep[see, e.g.,][]{Giles1996}. In our analysis, we exclude different burst phases from the light curves and concentrate only on the flat part, because we can assume that during that phase the conditions in the disk remain constant. 

The program \textsc{powspec} was used to calculate the PDS from the X-ray light curves. Discrete PDSs were calculated from 16~s long segments of the light curves and averaged together for each observation to get a PDS in a range of 0.0625$-$4096~Hz. 
The PDSs were further averaged within different luminosity levels as shown in Figure \ref{fig:lc_entire}. Again, we only considered the data taken during  main outbursts marked by O1 and O2.
The data files in each list given to \textsc{powspec} must have the same time resolution, therefore some light curves were rebinned to match the lowest time resolution for the selected set of light curves. The aforementioned single-bit configurations which were separated into two different energy bands (channel ranges 0--23 and 24--249) were combined to get the counts in the entire PCA energy range. 

The broadband PDSs were normalized to the  mean count rate squared, so that integration of the PDS gives the fractional rms variability \citep{Miyamoto1991}. 
The level of the Poisson noise was estimated by averaging the nearly flat high-frequency part from 1000 to 3000 Hz in each PDS and then subtracted. We do not see deadtime or instrumental effects affecting the Poisson noise in this frequency range.
The PDS from the lowest luminosity bins were excluded from the analysis due to poor statistics and possible background contamination.

\subsection{Spectral analysis}
\label{sec:spec}

To calculate the luminosities, we extracted also the energy spectra for all observations following standard procedures of {\it RXTE} Cook Book guide. In a similar way to the timing analysis, spectra were extracted only from the flat periods in the light curves. Python interface to \textsc{xspec}, \textsc{PyXspec}, was used for fitting the spectra between 2.5 and 45 keV, where the most of the flux is observed. As spectral model we consider a power law with an exponential high-energy cut-off and a Gaussian line, all absorbed by  interstellar matter \textsc{phabs*(powerlaw*highecut+gauss)}. The Gaussian component is added to account for the wide fluorescent iron K$\alpha$ line with central energy fixed to 6.7~keV and width to 0.3~keV \citep[see, e.g.,][]{Doroshenko2015}. The hydrogen column density was fixed to $6.2\powten{22}$~cm$^{-2}$, the average value for all the observations. 
This spectral model was used to estimate the absorption corrected flux with the errors corresponding to statistical uncertainties from the spectral fit. 
The  source luminosity was obtained assuming isotropic emission and the distance of 8~kpc (see  Figure \ref{fig:lc_entire}). 
 
\section{Results}
\label{sec:results}

\subsection{Shape of power density spectra}
\label{sec:pds_results}

\begin{figure}
\centering
\includegraphics[width=\columnwidth]{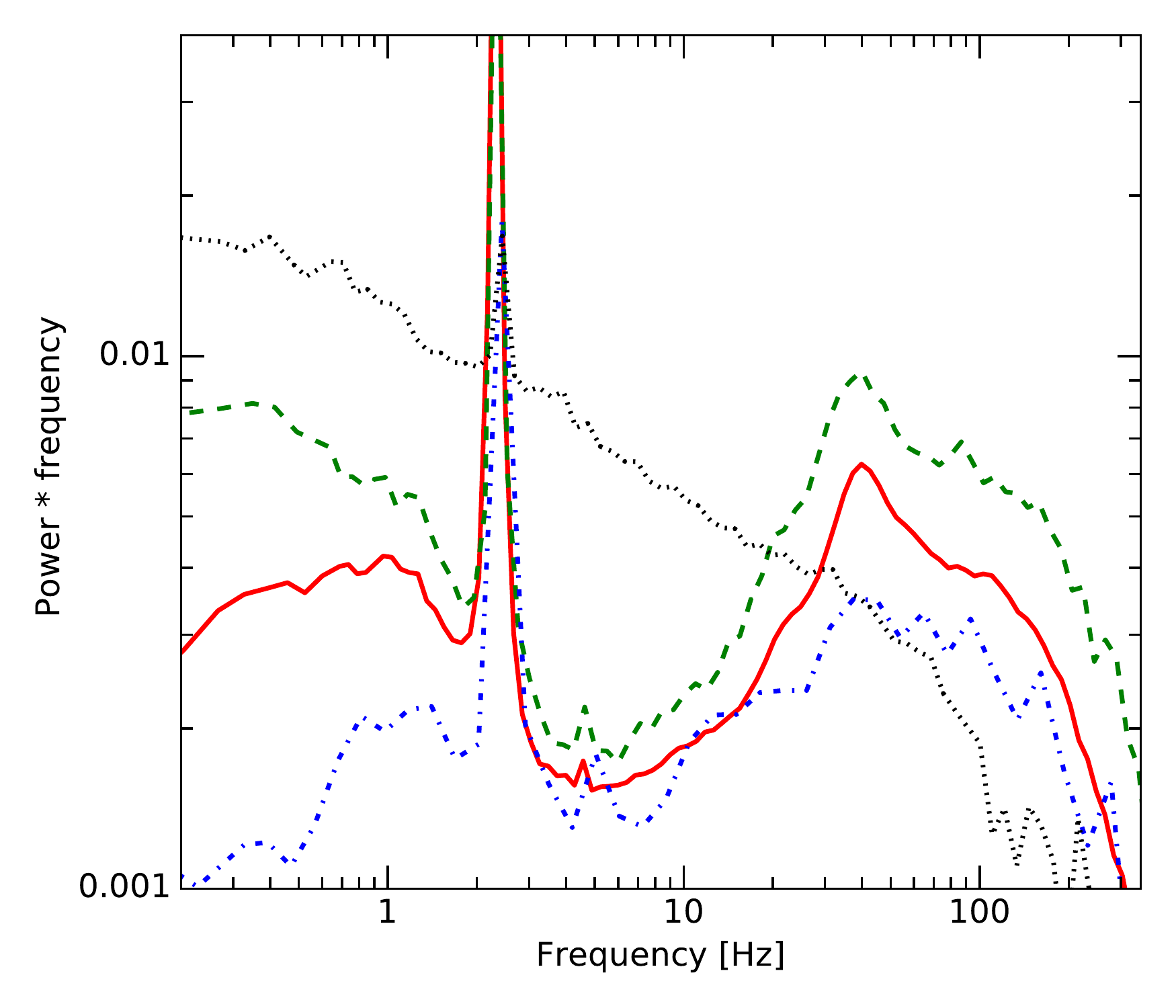}
\caption{Broadband PDS of the persistent light curve corresponding to different luminosity levels during the second outburst decay and the first mini-outburst. 
The red line corresponds to the luminosity of $L=5.0\powten{38}$~\lum, the green dashed line to $L=7.6\powten{37}$~\lum, blue dot dashed to $L=2.5\powten{37}$~\lum. The dotted line is from the mini-outburst at luminosity $L=8.8\powten{37}$~\lum. The PDS were logarithmically rebinned with respect to frequency to suppress the noise at the highest frequencies. The dominating regular pulsations are visible around 2 Hz. QPO features are seen around 20, 40 and 60~Hz. A PDS component between 5 and 200 Hz is referred to as the high-frequency bump. Note that Miyamoto normalization was used, the rebinning at lower luminosity seemingly shifts the pulse frequency and error bars were left out for clarity.
}\label{fig:pds}
\end{figure}

\begin{figure}
\centering
\includegraphics[width=\columnwidth]{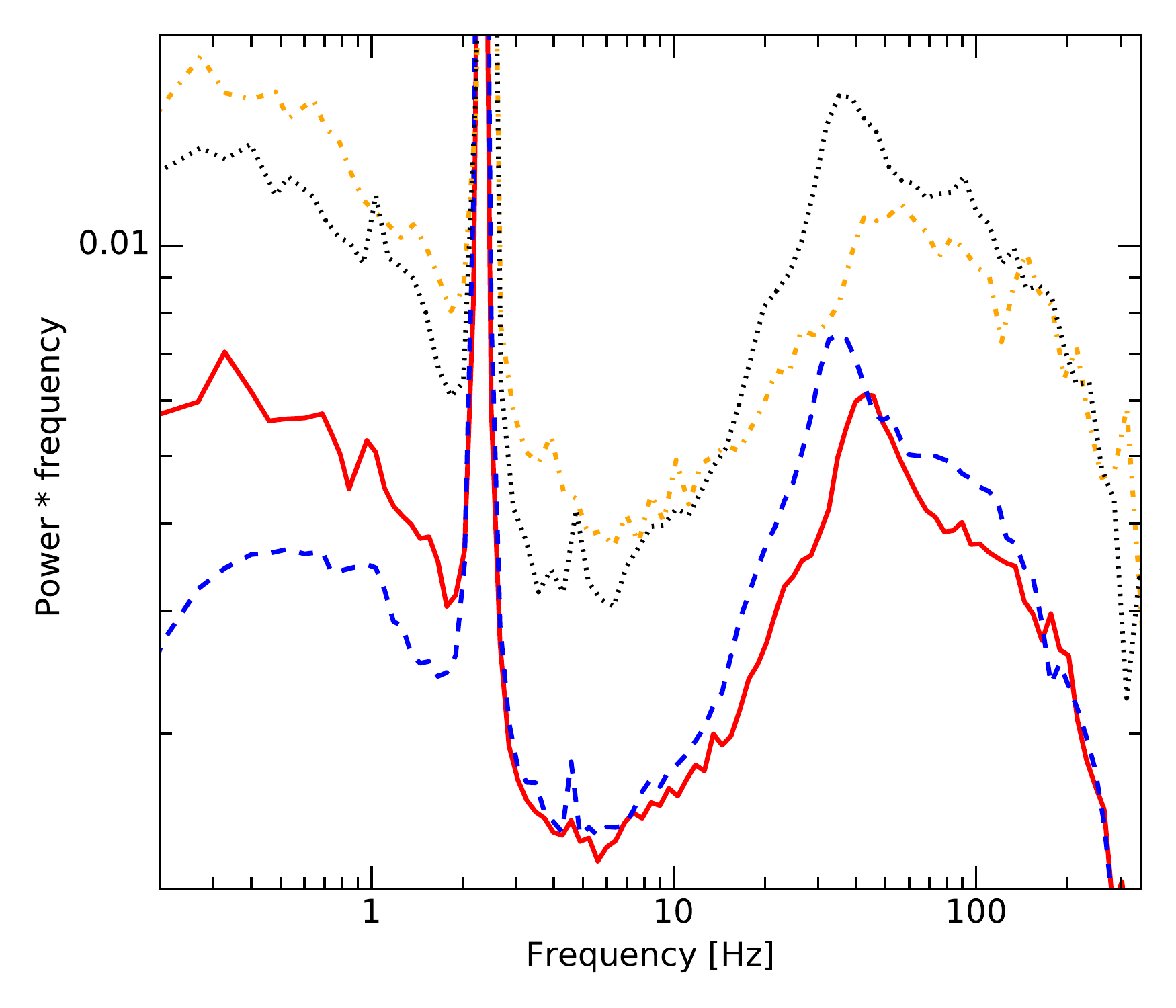}
\caption{A comparison of broadband PDS of \gro\ from both of the outbursts at two different luminosity levels where the measured break frequencies showed large discrepancies. The red solid line is from outburst 1 at $L=(2.84\pm0.02)\powten{38}$~\lum, the blue dashed line from outburst 2 at $L=(3.0\pm0.3)\powten{38}$~\lum, the orange dot-dashed line from outburst 1 at $L=(1.2\pm0.2)\powten{38}$~\lum\ and the black dotted line from outburst 2 at $L=(1.3\pm0.2)\powten{38}$~\lum. The PDS at the lower luminosities are multiplied by factor of 2 for better visualization. }\label{fig:pds_discrepancies}
\end{figure}

Examples of the PDS at different luminosities are shown in Figure \ref{fig:pds}. In the major outbursts of \gro, the shape of the PDS differs from the standard broken power-law expected for XRPs with gas-pressure dominated (GPD) accretion disks. Instead, two bumps in ranges 0.1$-$5~Hz and 5$-$200~Hz are observed. The high-frequency bump is shaped approximately as a broken power-law with indices $-0.5$ and $-2.5$ and a break around 100~Hz. The narrow peaks corresponding to the nearly sinusoidal, regular pulsations at 2.14 Hz and their weaker second harmonic are seen right in the valley between the bumps. Quasi-periodic oscillations (QPOs) are present around 20, 40 and 60 Hz \citep[see][]{Zhang1996, Kommers1997}. The shape of the PDS evolves with the luminosity during the outburst decay phase. In excluded mini-outbursts occurring after the major outbursts, the shape of the PDS (dotted black line in Figure \ref{fig:pds}) has changed dramatically, and can be characterized by a broken power law similar to that observed in other XRPs. The break in mini-outbursts appears at frequencies nearly as high as in the major outbursts, which suggests that in all of the outbursts, the break in PDS has a common origin. There are also slight differences between the major outbursts. We show a comparison of the corresponding PDSs in Figure~\ref{fig:pds_discrepancies} at two luminosity levels, $L\approx 3\times 10^{38}$ and $\approx1.3\times 10^{38}$~\lum, where the measured break frequency shows large discrepancy between the outbursts (see the next section). 

\begin{table*}
    \centering
    \caption{Parameters of the fits to the PDSs.}
    \begin{tabular}{ccccc}
    \hline
        \hline
        $L$ & $f_\text{b}$  & $f_\text{QPO,1}$ & $f_\text{QPO,2}$  & $f_\text{QPO,3}$  \\
         ($10^{38}$ erg s$^{-1}$) & (Hz) & (Hz)  & (Hz)  & (Hz)   \\        
        \hline
        \multicolumn{5}{c}{ Outburst 1}\\
        $0.74\pm0.09$ & $130\pm30$ & $18\pm2$  & $37\pm2$  & $53\pm2$ \\
        $1.2\pm0.2$ & $140\pm20$ & $23.2\pm0.7$  & $40\pm2$  & $55\pm3$ \\
        $1.8\pm0.3$ & $120\pm20$ & $22.5\pm0.6$  & $39\pm4$  & $47\pm4$ \\
        $2.84\pm0.02$ & $130\pm6$ & $21.9\pm0.4$  & $38.3\pm0.4$  & $52\pm2$ \\
        $5.0\pm0.9$ & $130\pm4$ & $19.6\pm0.3$  & $35.1\pm0.2$  & $52\pm1$ \\
    $8.1\pm0.9$ & $141\pm4$ & $20.2\pm0.3$  & $36.1\pm0.2$  & $56\pm1$ \\
        \hline
        \multicolumn{5}{c}{ Outburst 2}\\
        $0.76\pm0.08$ & $100\pm10$ & $17.0\pm0.9$  & $30.5\pm0.5$  & $45\pm5$ \\
        $1.3\pm0.2$ & $93\pm9$ & $18.1\pm0.5$  & $31.7\pm0.6$  & $45\pm9$ \\
        $1.9\pm0.3$ & $111\pm5$ & $20.5\pm0.4$  & $33.5\pm0.2$  & $49\pm2$ \\
        $3.0\pm0.3$ & $100\pm3$ & $18.0\pm0.3$  & $31.3\pm0.2$  & $49.0\pm0.8$ \\
        $5.0\pm0.7$ & $132\pm2$ & $20.2\pm0.2$  & $36.0\pm0.1$  & $55.7\pm0.5$ \\
\hline        
    \end{tabular}
    \label{tab:pds_params_1}
\end{table*}

\subsection{Break frequency}
\label{sec:breaks}

The PDSs of \gro\ have varying shapes but the high-frequency bump can be fit with a smoothly broken power law  
\be 
\label{eq:bpl}
P(f)= N \frac{(f/f_{\text{b}})^a}{[(f/f_{\text{b}})^{d(a-b)}+1]^{1\! /d}},
\ee
where $f$ is the frequency variable, $f_{\text{b}}$ is the break frequency, $a$ is the exponent of the first slope, $b$ is the exponent of the second slope, $N$ is a constant. 
The smoothness of the transition  is defined by parameter $d$ \citep[see][]{Revnivtsev2009, Revnivtsev2010} and was fixed at 2 in order to standardize how the break frequency corresponds to the transition from one slope to another; a higher value of $d$ decreases the measured value of the break frequency and vice versa. In the outbursts, three Lorentzians were included in the model to account for the QPOs:   
\be
\label{eq:qpo}
P_{\text{QPO}}(f) = \frac{N_\text{QPO}}{(f-f_{\text{QPO}})^2+(w/2)^2},
\ee
where $N_\text{QPO}$ is a constant, $f_{\text{QPO}}$ is the QPO centroid frequency and $w$ is its width. 

\begin{figure}
\centering
\includegraphics[width=\columnwidth]{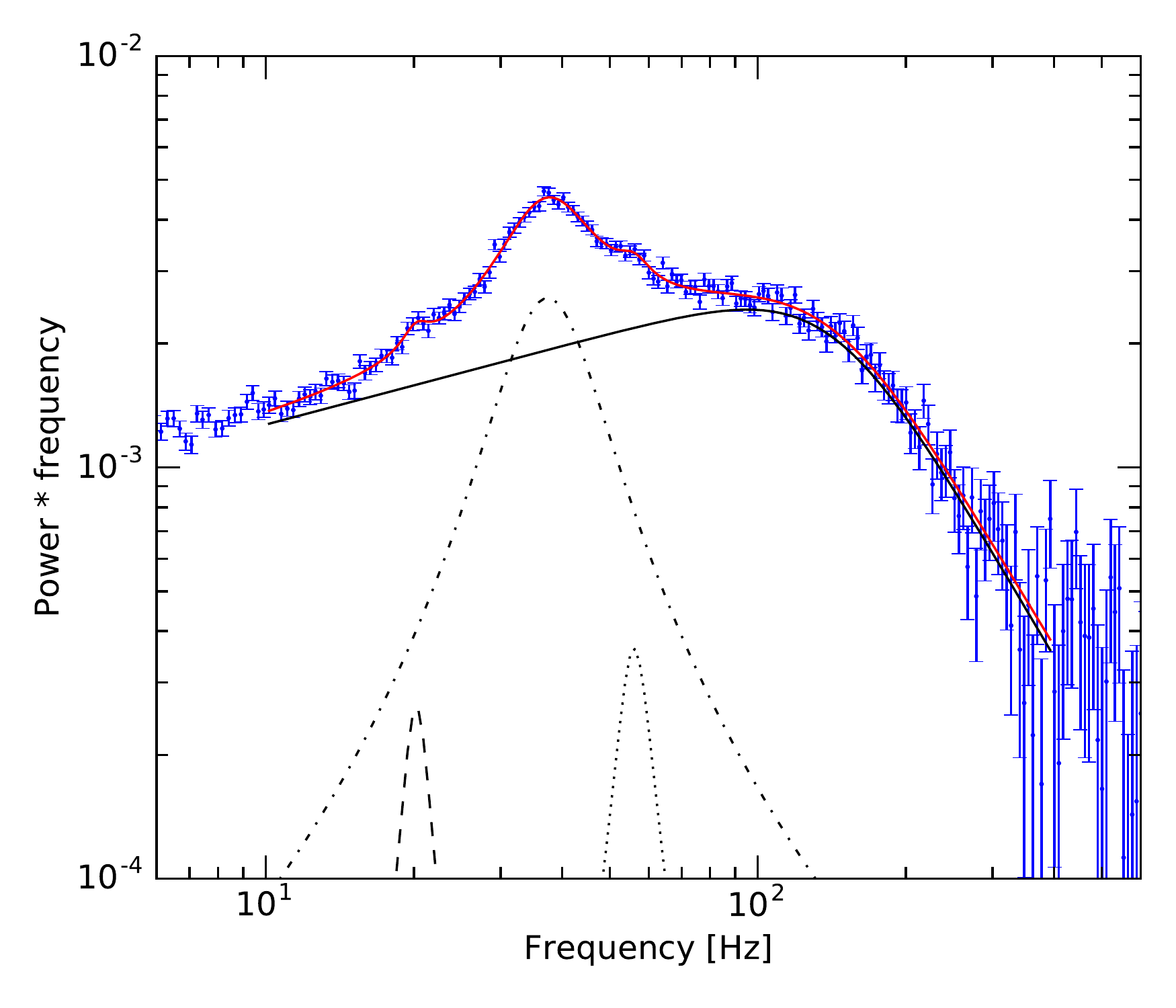}
\caption{The observed PDS of \gro\  at the peak of the first outburst (blue dots with error bars) and the best-fit model (red curve) in a frequency range from 10 to 400 Hz. 
The model consists of  a smoothly broken power law (solid black curve) and  three Lorentzians shown by dashed, dot-dashed and dotted curves. }\label{fig:fit_pds}
\end{figure}

The fits to the observed PDSs were conducted with the \textsc{curve\_fit } function (non-linear least squares) of  \textsc{scipy.optimize} package for \textsc{Python }over a frequency range from 10 to 400 Hz (see an example in Figure \ref{fig:fit_pds}). Because many PDSs were averaged together, variation in the break frequency with luminosity may have smoothened the transition. The power-law index of the second slope also increases with luminosity, but our modeling showed that this does not affect the break frequency measurement.

Break frequencies and QPO centroid frequencies for the outbursts are shown in Table \ref{tab:pds_params_1}. The break frequencies vary approximately between 90 and 140 Hz in the six highest luminosity bins during the outbursts. The QPO centroid frequency, on the other hand, remains stable with luminosity although in the second outburst the values seem to be somewhat lower. One should note that the power spectra are calculated from several observations which might increase the QPO width and affect the determination of the centroid frequencies.

\begin{figure}
\centering
\includegraphics[width=\columnwidth]{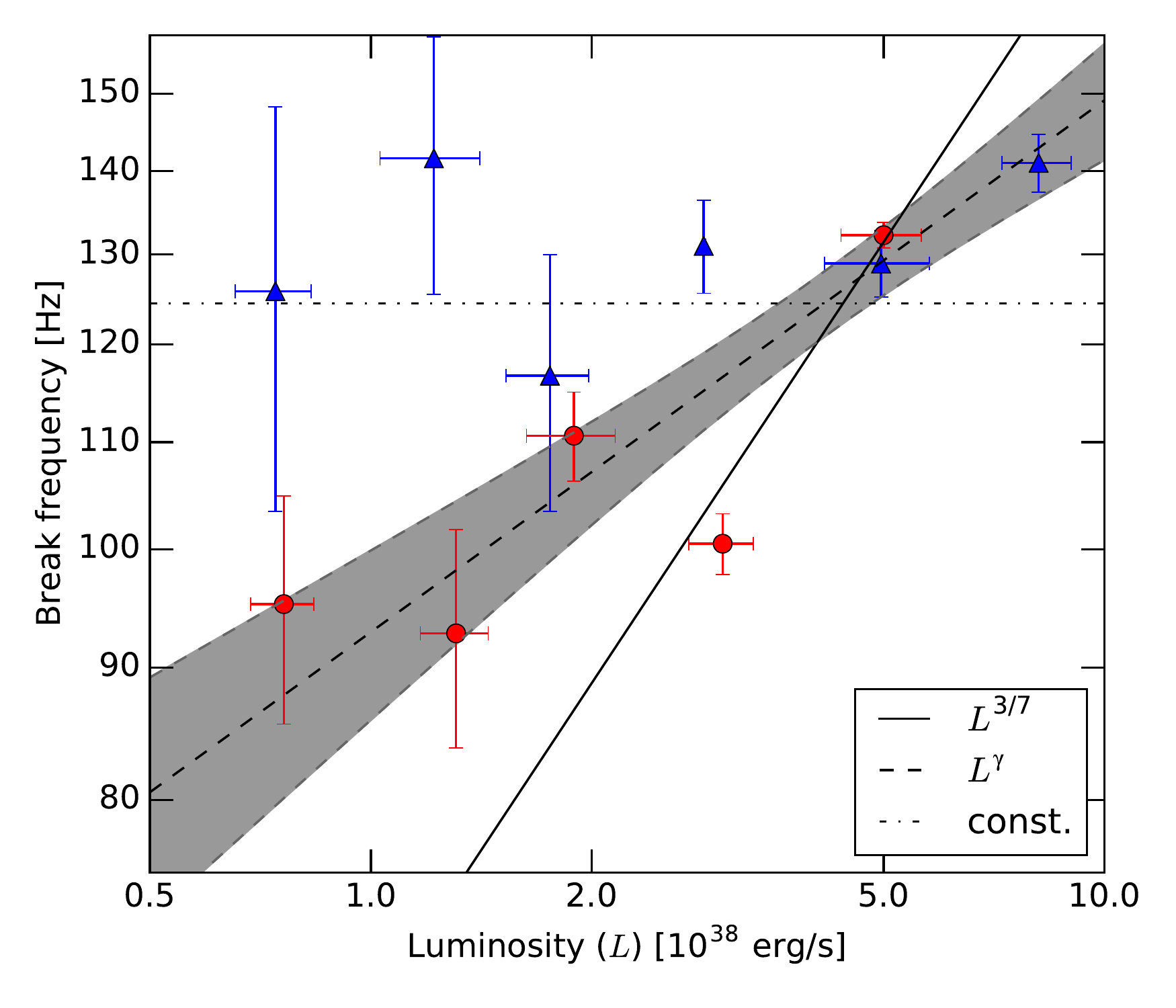}
\caption{Break frequencies from the fits to the high-frequency bumps of PDSs plotted against the luminosity. Blue triangles correspond to the first major outburst and red circles to the second major outburst. The fits of luminosity-dependencies to break frequency data from the major outbursts are shown. The scaling constant is a free parameter in all of the fits. For the free slope, we get an index of $\gamma=0.21\pm 0.02$, and the confidence interval calculated from the fit scatter is also provided (shaded gray).
}\label{fig:lumin_fbreak}
\end{figure}

The measured break frequencies as a function of luminosity are presented in Figure \ref{fig:lumin_fbreak}. The luminosity values are the averages of all the individual observations in each luminosity range and the errors are their standard deviation (better describing the distribution of luminosities of individual observations within the luminosity range than just the range limits). Some data points at lowest luminosities were left out due to poor statistics. We examined the luminosity-dependence of the break frequency by fitting the break frequency data from both of the major outbursts when the behavior is expected to be similar. We used a power-law relation $f_\text{b}=CL^\gamma$ with three different indices: (i) $\gamma$ fixed to $3/7$ (resulting from Eq.~\eqref{eq:kepl} by substituting $R_\text{m}$ given by Eq.~\eqref{eq:r_m}), (ii) $\gamma$ fixed to $0$ \citep{Chashkina2017} and (iii) $\gamma$ being free to vary. Fitting was done using the Monte Carlo method: the data points were let to vary within the given errors in both frequency and luminosity. The slopes which were calculated without taking into account the errors in luminosity give almost the same results, and therefore, the $\chi^2$ could be calculated by using only the errors in the break frequency.
The reduced $\chi^2$ for the fits with fixed exponents $3/7$ and $0$ were $15$ and $16$, respectively. 

Letting the slope to vary, we got $\gamma = 0.21\pm 0.02$ with still a rather high reduced $\chi^2=6.8$ due to the scatter of the data points. Therefore, we also fit the break frequencies of both outbursts separately. For the first outburst, we got $\gamma = 0.06\pm 0.02$ with $\chi^2\approx 1.1$, and for the second outburst, $\gamma = 0.28\pm 0.03$ with $\chi^2\approx 11$ (due to a low number of scattered data points). Possible physical explanations for the luminosity-dependence are discussed in Sect.~\ref{sec:depend}

\section{Discussion}
\label{sec:discuss}

\subsection{Inner disk radius in \gro}
\label{sec:inner}

The main result of the timing analysis presented above, is a rather unusual PDS which deviates strongly from that observed in normal XRPs. It is important to emphasize that the presence of high-frequency variability in the Bursting Pulsar suggests that the accretion disk extends close to the NS because no additional noise is expected to be generated at frequencies higher than the Keplerian frequency at the innermost disk radius. Our discovery of the break in the high-frequency range in both the mini-outbursts and the major outbursts can thus be connected to the inner radius of the disk. In the perturbation propagation model, the observed variability can be linked to the orbital velocities $\omega$ of the disk matter. Down to the magnetospheric boundary, the orbital velocities as a function of disk radius $R$ are Keplerian:
\be
\label{eq:kepl}
f(R)= \omega(R)/2\pi = \frac{1}{2\pi} \sqrt{\frac{GM}{R^3}},
\ee
where $G$ is the gravitational constant and $M$ is the NS mass. Following \citet{Revnivtsev2009}, we associate the break frequency with the maximal frequency of disk variability, that is, the Keplerian frequency at the innermost disk radius. Solving Eq.~\eqref{eq:kepl} we get for the radius 
\be
\label{eq:r_f}
R_\text{m}(f_\text{b}) = \left(\frac{GM}{(2\pi f_\text{b})^2}\right)^{1/3}. 
\ee 
For \gro, substituting the highest measured break frequency of $141\pm4$~Hz to Eq.~\eqref{eq:r_f} gives a value for the inner radius of only $(6.2\pm0.2)\powten{6}$~cm at luminosity of $8\powten{38}$~\lum. During the peak of the outburst, the inner radius would then be only 4--6 NS radii for the assumed $M=1.4M_\odot$. 

On the other hand, it is possible to estimate the magnetospheric size $R\subt{m}$ using the magnetic field strength estimated from the cyclotron line, assuming a pure dipole field. Equalizing the magnetic pressure with the ram pressure of matter in a free fall we get \citep{Lamb1973,King2002},
\be
\label{eq:r_m}
R\subt{m}= 2.6\powten{8}\ \xi \ m^{1/7} R_{*,6}^{10/7} B_{12}^{4/7} L_{37}^{-2/7} \text{cm},
\ee
where $m=M/M_\odot$ is the NS mass in units of solar masses, $R_{*,6}$ is the radius of the NS in $10^6$~cm, $B_{12}$ is the magnetic field strength in $10^{12}$~G and $L_{37}$ is the luminosity in $10^{37}$~\lum. Here it was assumed that the observed luminosity is proportional to the mass accretion rate. 
An additional parameter $\xi$ accounts for the deviation of the accretion flow  geometry from spherical. 
As mass accretion rate increases, the disk is able to push itself further into the magnetosphere and for a GPD inner disk  the inner radius decreases as $R\subt{m}\propto L^\beta$ with $\beta=-2/7\approx -0.29$. 
The  factor, $\xi$, is typically assumed to be 0.5$-$1 \citep[however, see also][]{Kulkarni2013,Chashkina2017,Chashkina2019}. In Bursting Pulsar, we can compare the inner radius calculated from the break frequency using Eq.~\eqref{eq:r_f} to the magnetospheric radius with Eq.~\eqref{eq:r_m}. We fix the magnetic field $B_\text{cycl}= 5\powten{11}$~G and solve for $\xi$. For the $R_\text{m}=6.2\powten{6}$~cm calculated above at $L_{37}=80$, we get $\xi\approx 0.09$, assuming $m=1.4$ and $ R_{*,6}=1.2$ \citep{Suleimanov2017}. The required value of $\xi$ is much smaller than obtained by simulations and analytic models.

We note that the equation for the magnetospheric radius given in Eq.~\eqref{eq:r_m} is a simplification. An additional parameter which affects the inner radius, is the inclination of the NS magnetic axis with respect to the disk. \cite{Bozzo2018} examined this effect and found that the inner radius for an inclined dipole is smaller. All in all, the radius can vary approximately by a factor of two between the extreme cases. Therefore, inclination by itself is not enough to explain the discrepancy in the measured radius values, especially considering the limitations for the inclination because the observed flux comes entirely from one magnetic pole \citep{Strickman1996}. 

The inner radius of the accretion disk of \gro\ has been estimated previously by several authors based on different consideration, and, in fact, the radius was often found to be smaller than expected from the magnetic field producing the cyclotron line. \citet{Nishiuchi1999} based on the modeling of a broad iron line observed by \textit{ASCA} satellite, obtained an estimate of  $(4-12)\powten{6}$~cm for the inner radius. More recently, \citet{DAi2015} modeled \textit{XMM-Newton} and \textit{INTEGRAL} broadband continuum spectra, assuming that soft excess at $kT\subt{disk}=0.54$~keV corresponds to the maximal temperature of a multicolored disk, and got $R\subt{in}=(1.1_{-0.3}^{+0.8})\powten{7}$~cm near the peak of the outburst (at a luminosity of $2.1\powten{38}$~\lum\ for a distance of 8~kpc). The disk model normalization, on the other hand, yielded a somewhat smaller radius of $6\times 10^6$~cm which matches our estimates rather well. \citet{DAi2015} acknowledged that these estimates imply rather small inner disk radii and concluded that $\xi\approx 0.2$. \citet{Younes2015} used data from \textit{Chandra} and \textit{NuSTAR} right after the peak of the 2014 outburst to study the reprocessing region of the disk, and estimated inner radius of $R\subt{in}= (4\pm 1)\powten{7}$~cm from the blackbody component with $kT\approx 0.6$~keV. This result was consistent with their analysis of the $6.7$~keV broadened iron line. Finally, \citet{Degenaar2014} also studied \textit{Chandra} data and used the same $6.7$~keV iron feature to derive a similar inner radius of $2\powten{7}$~cm. Therefore, although there is a significant dispersion in measured values of inner disk radius in \gro, all estimates tend to yield a relatively small radius and the lower range of estimates is consistent with our results and at odds with the magnetospheric radius estimates using the observed cyclotron line energy.

One possible explanation for this discrepancy is that the magnetic field does not have a pure dipole configuration. Instead, it could be deformed due to interaction with the disk, or there could be multipole components in the NS magnetic field, affecting the field configuration and the cyclotron line energy. The multipole moments will decay faster with respect to distance than the pure dipole field: for example a quadrupolar field has $B\propto R^{-4}$ \citep[e.g.][]{Barnard1982}. In this case, we could have a situation where the cyclotron line is produced close to the NS in a stronger magnetic field affected by the multipoles, while the disk only feels the weaker dipole field away from the NS, allowing the matter to come closer to the NS. 
We note that \citet{Tsygankov_smcx3} invoked the same argument to explain the small magnetosphere size in ultra-luminous X-ray pulsar SMC~X-3. 

We can estimate the magnetospheric radius in case of a purely quadrupolar magnetic field. We do not take into account the 3-dimensional shape of the quadrupolar field but simplistically take $\mu_{\text{q}}=B_{\text{q}}R^4$ following \citet{King2002} (p. 158) and arrive at
\begin{equation}
\label{eq:r_m_q}
    R_{\text{m,q}} = 3.7\times 10^7 \xi_\text{q} m^{1/11} R_{*,6}^{14/11} B_{12}^{4/11} L_{37}^{-2/11}\ \text{cm.}
\end{equation} 
In the equation, the $\xi_\text{q}$ parameter again measures the deviation from spherical accretion and we also note that it may be different from the dipole case because the quadrupolar magnetosphere probably resists deformation more than the dipole. For the Bursting Pulsar at $L=8\powten{38}$~\lum, taking $\xi_\text{q}=1$, we get  $R_{\text{m,q}}=1.7\times 10^7$~cm which corresponds to a Keplerian frequency of $\sim$30~Hz that is still a factor of a few lower than the observed one. Hence, $\xi_\text{q}$ as small as $\approx 0.4$ is still required to be consistent with the data. 

Alternatively, one can assume that the effective magnetospheric radius is indeed small compared to Alfv\'enic radius. Indeed, there is a large uncertainty in value of the coupling constant $\xi$ resulting from uncertainty in the accretion disk structure. The fact that the observed PDS shape deviates from the broken power law commonly observed in other XRPs, strongly suggests that the structure of accretion disk inner regions might indeed be different in \gro. This scenario is discussed in detail below.

\subsection{Disk-magnetosphere interface}
\label{sec:disk}

In this section, we will discuss the structure of the accretion disk and some of the consequences of high luminosity in \gro.
First of all, it is important to emphasize that the observed high frequency variability implies that the disk extends very close to NS surface, which is to be expected given the moderate field of the source.
This, together with the observed
high luminosity of the source, implies that the inner edge of the disk must be in the zone A of the standard Shakura-Sunyaev accretion disk \citep{Shakura1973}.
Thus the inner regions would be radiation-pressure dominated (RPD) instead of GPD, which can be
expected also to have some impact on the observed break frequency and PDS shape in general. The difference in the PDS shape between the mini-outburst and the major outburst is perhaps an indication of such a change in the physical properties of the system.

Transition from zone C (opacity dominated by absorption) to B (electron scattering dominating region) happens at a radius
\be
\label{eq:r_bc}
R_\text{BC} = 5\powten{7} m^{1/3} \dot{M}_{17}^{2/3}\ \text{cm},
\ee
where $\dot{M}_{17}$ is the mass accretion rate in units of $10^{17}$~g~s$^{-1}$. Here solar chemical composition is assumed and the opacities are  taken to be independent of the height with their characteristic values in the disk \citep{Suleimanov2007}. The boundary between GPD and RPD zones can be estimated from the balance between gas and radiation pressures in the disk:
\be
\label{eq:r_ab}
R_\text{AB} = 10^7 m^{1/3} \dot{M}_{17}^{16/21} \alpha^{2/21}\ \text{cm},
\ee
where $\alpha \lesssim 1$ is the viscosity parameter of a Shakura-Sunyaev disk, and as in Eq.~\eqref{eq:r_bc}. 
Some uncertainty is related to this estimate because of the uncertainty in the chemical abundance.
\citet{Rappaport1997} showed in their evolutionary modeling that it is also possible for the companion to be a small helium-rich star, which would mean a helium-rich composition of the disk \citep[cf.][]{Gosling2007}. However, this composition only reduces the boundary radius of zone A by 15\%, and such as change is not large enough to affect our analysis. 
 
The threshold luminosity for the appearance of zone A, $L_\text{A}$, can be estimated from the condition that the inner radius $R_{\rm m}$ given by Eq.~\eqref{eq:r_m} equals  $R_\text{AB}$ \citep[see also][]{Andersson2005}:
\be
L_\text{A} = 3\powten{38} \xi^{21/22} \alpha^{-1/11}  m^{6/11} R_{\text{*,6}}^{7/11} B_{12}^{6/11}\ \text{erg s$^{-1}$}.
\ee
In the case of \gro, we get an estimate for the threshold to be $L_{\text{A}} = 1.8\powten{38}$~\lum\ for $B=5\times 10^{11}$~G, $\alpha=0.1$ and $\xi=0.5$. One should note that the threshold depends strongly on the value of $\xi$, so for a smaller $\xi=0.1$, we get $L_{\text{A}} = 0.4\powten{38}$~\lum. We have also assumed the NS mass of $1.4M_\odot$ and radius of 12~km  \citep{Steiner2016EPJA,Lattimer2016PhR,Nattila2017}. 
In both of the outbursts we examined, the source luminosity reaches values which are several times above the threshold luminosity. Consequently, we expect the accretion disk in \gro\ to have a RPD inner region. 

\begin{figure}
\centering
\includegraphics[width=\columnwidth]{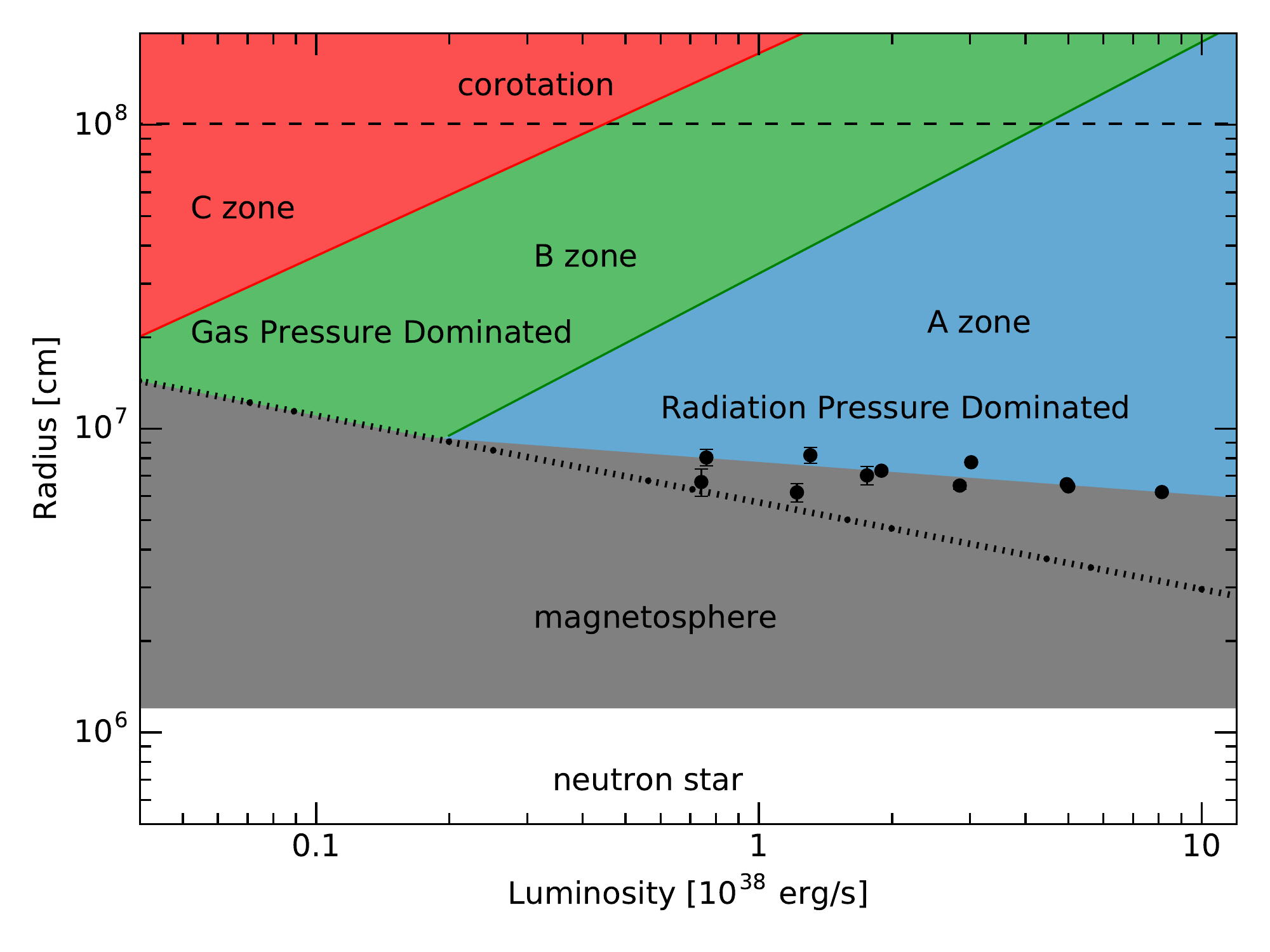}
\caption{Schematic slices of the accretion disk in radial direction at different accretion luminosities. Different zones are named and the corotation radius  for \gro\ is shown with the dashed line. Here we assume $m=1.4$, $R_{*,6}=1.2$, $\alpha=0.1$ and $B=5\times 10^{11}$~G. 
The observed magnetospheric truncation radii in the RPD region for \gro\ are shown with the black circles (see Sections \ref{sec:breaks} and \ref{sec:inner}). 
The truncation radius depends on luminosity as $\propto L^{-0.1}$ (the boundary between blue and gray regions). The dotted line shows the model truncation radius computed using Eq.~\eqref{eq:r_m}. The parameter $\xi$ is taken to be 0.05 to smoothen the transition from the observed truncation radius in the RPD region. }  
\label{fig:disk_zones}
\end{figure}

The estimates for the boundary radii between zones A, B and C, which  are independent of the magnetic field strength, are illustrated  in Figure \ref{fig:disk_zones} for different luminosities. We also show the magnetospheric truncation radius which is estimated for the Bursting Pulsar by assuming that the observed break frequencies are the Keplerian frequencies of the inner radius.
With this assumption, we see that the estimated inner radius is in the zone A. The luminosity-dependence of the inner radius is not based on theoretical models, but rather, a free power-law fit to the data points, resulting in a slope of $-0.1$. When we extrapolate the estimated inner radius of the RPD region to lower luminosities with the observed luminosity-dependence, we see that the zone A would exist in the disk above a threshold of $L\approx 2\powten{37}$~\lum\ for $\alpha=0.1$. The observed threshold luminosity is low and suggests that either the $\xi$ parameter must be much lower than 0.5 or that the magnetic field does not behave as expected. The shape of the PDS in major outbursts supports the presence of a RPD region at least down to these luminosities. Interestingly, in the decay phases of the major outbursts of \gro, the bursting halts when the luminosity approaches this observed threshold.

Indeed, \citet{Court2018a} examine the bursting behavior and find that the typical type II bursts (similar to those seen in Rapid Burster) only appear above a threshold of $0.1L_\text{Edd}$ corresponding to $2\powten{37}$~\lum\ for a distance of 8~kpc, while below this threshold, the bursting behavior is different. This luminosity is in good agreement with our observational estimate of the threshold for the occurrence of the RPD inner disk region. \citet{Ji2019} analyzed the spectral hardness-flux diagram and the hard X-ray lags and found that there is a transition in their properties at luminosity of $8\powten{37}$~\lum, which they associate with a change of the accretion geometry. We note that below that luminosity, the post-burst dips become  shallower compared to the dips at the peak of outbursts, probably indicating that the size of the burst reservoir has decreased.  

 \subsection{Modeling PDS of a disk with a RPD inner region}
\label{sec:model_pds}

According to the model of propagating fluctuations of the mass accretion rate \citep{Lyubarskii1997, Kotov2001, Churazov2001, Arevalo2006, Ingram2013, Mushtukov2018,2019MNRAS.tmp..913M}, the disk produces the initial fluctuations, while the viscous properties of the disk determine the propagation process and the final properties of aperiodic variability over the entire accretion flow.
The diffusion process effectively suppresses high frequency variability and, as a result, each radial coordinate in the disk contributes effectively only to variability below local viscous frequency. 
The viscous properties of the disk are described by the kinematic viscosity $\nu$, which is assumed to be a function of radial coordinate only. 
This assumption is justified because specially for $\alpha$ disks of constant $\alpha$-parameter \citep{Blaes2014SSRv}, where gas pressure and Kramer opacity dominate (C-zone), the scale-height is $H/R\propto R^{1/8}$ and therefore the kinematic viscosity is a function of radius only $\nu = \alpha \Omega_{\rm K} R^2 (H/R)^2 \propto R^{3/4}$ \citep{Shakura1973,Mushtukov2018}. 

At high mass accretion rates, however, the inner part of accretion disk becomes RPD. 
The dependence of the thickness of the accretion disk on the radial coordinate is different from that in the GPD part,
with the thickness being almost independent of radius \citep{Shakura1973}. 
Thus, the dependences of the kinematic viscosity and viscous time scale on radius become different. 
In particular, the viscous time scale becomes shorter and viscous diffusion is expected to suppress less effectively the initial variability at high frequencies.
Therefore, one would expect an appearance of an excess of aperiodic variability power at high frequencies in accretion disks with RPD inner parts.\footnote{We have to notice that the question about stability of RPD part of accretion disk is still open. The standard RPD disks have been shown to be unstable \citep{Shakura1976}, but the instability predicted in theory has never been detected \citep{Done2007}.  At the same time, it has been shown that the aperiodic variability by itself can principally stabilize RPD accretion flow  \citep{Sukova2016}. 
See also \citet{Cannizzo1996, Cannizzo1997} discussing the bursts of \gro.}

In order to illustrate the effect arising from the appearance of a region with higher viscosity in accretion disk, we have to investigate the process behind the formation of PDS.
The process of matter diffusion along the radial coordinate $R$ in accretion disk is described by the viscous diffusion equation 
\be
\label{eq:diffusion}
\frac{\partial \Sigma(R,t)}{\partial t} = \frac{1}{R}\frac{\partial}{\partial R} \left[ R^{1/2} \frac{\partial}{\partial R} \left( 3\nu \Sigma R^{1/2} \right) \right],
\ee
where $\Sigma$ is the local surface density in accretion disk \citep{Pringle1981}. 
If the kinematic viscosity $\nu$ is a function of radius only,  Eq.~\eqref{eq:diffusion} is linear and its solutions can be expressed through the Green's function $G(R,R',t)$. They depend on the viscosity $\nu$ and the boundary conditions at the inner and the outer disk radii, $R_\text{in}$ and $R_\text{out}$.

PDS of the mass accretion rate is determined by aperiodic variability arriving to a given radius $R$ from all other radial coordinates $R'$ \citep{Mushtukov2018} 
\be
P_{\dot{M}}(R,f) \simeq \int_{R_\text{in}}^{R_\text{out}} \frac{\text{d}R'}{(R')^2} \Delta R(R')\  \vert \bar{G}_{\dot{M}}(R,R',f)\vert ^2\  P_a(R',f),
\ee
where $f$ is the Fourier frequency, $P_{\dot{M}}(R,f)$ and $P_a(R,f)$ are the PDSs of the mass accretion rate variability and the initial perturbations at radius $R$,
$\Delta R(R')$ is the radial scale where the initial perturbations can be considered as coherent \citep[close to the local disk scale height $H$, see][]{Hogg2016}, 
$\bar{G}_{\dot{M}}(R,R',f)$ is the Fourier transform of the mass accretion rate Green's function of the viscous diffusion equation \citep[see e.g.][]{Kotov2001,Mushtukov2018,2019MNRAS.tmp..913M}.
The absolute value of the Fourier transform of the Green's function contains information about suppression of variability at different frequencies. 

For illustration, it is interesting to consider two accretion disk models with similar boundary conditions at the inner and outer radii, but different dependence of the kinematic viscosity on the radial coordinate: 
(a) $\nu\propto R^{3/4}$, which is typical for the C-zone in $\alpha$-disks, and 
(b) accretion disk with the inner part of higher kinematic viscosity: $\nu={\rm constant}$ at $R<5R_{\rm in}$ and $\nu\propto R^{3/4}$ at $R\geq 5 R_{\rm  in}$.

\begin{figure}
\centering
\includegraphics[width=9.3cm]{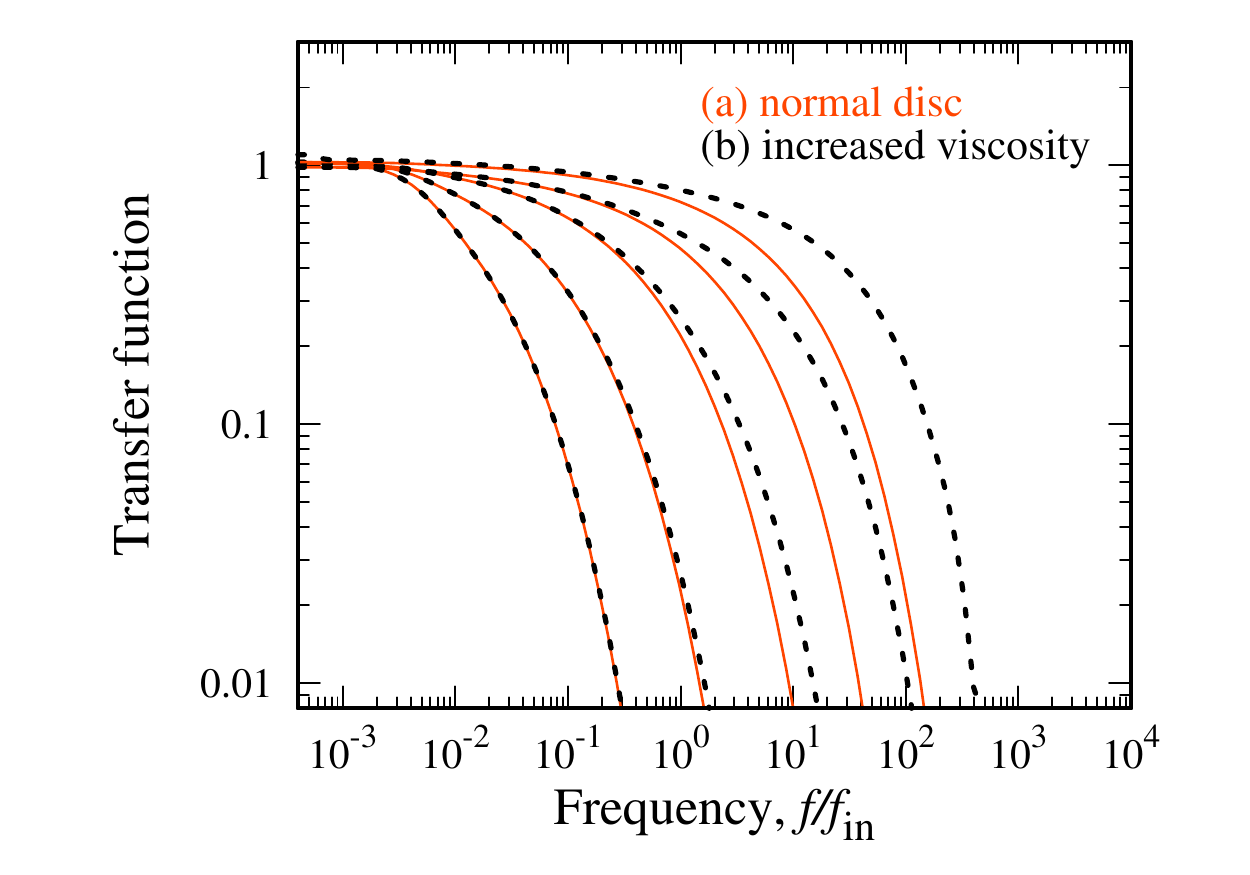}
\caption{The absolute value of the mass accretion rate Green's functions in the frequency domain calculated 
(a) for the case of GPD $\alpha$-disk (red solid curves), and 
(b) for the case of accretion disk with increased kinematic viscosity at the inner disk parts (black dashed curves).    
Different curves are given for various radial coordinates of initial perturbation: $R'=20,\,10,\,5,\,3,\,2R_{\rm in}$ (from left to the right). 
All Green's function are calculated for $R=R_{\rm in}$, i.e. describe suppression of variability at the inner radius of the accretion disk. 
Frequencies are normalized to the viscous frequency at the inner radius.
}\label{fig:sc_J04}
\end{figure}

Solving numerically the equation of viscous diffusion (\ref{eq:diffusion}), where the initial distribution of the surface density is given by the $\delta$-function, we get the Green's functions for the mass accretion rate and their Fourier transform (see Figure \ref{fig:sc_J04}). 
We see that the suppression of variability at high frequencies is indeed less effective in the case of accretion disk with the inner part of higher kinematic viscosity. 
The difference in suppression is more evident if the initial variability originates from the region of increased viscosity. 
As a result, we already can conclude that increase of kinematic viscosity in the inner parts of accretion disk can lead to less effective suppression of the high-frequency variability and, therefore, increased aperiodic variability at high frequencies. 

\begin{figure}
\centering
\includegraphics[width=9.3cm]{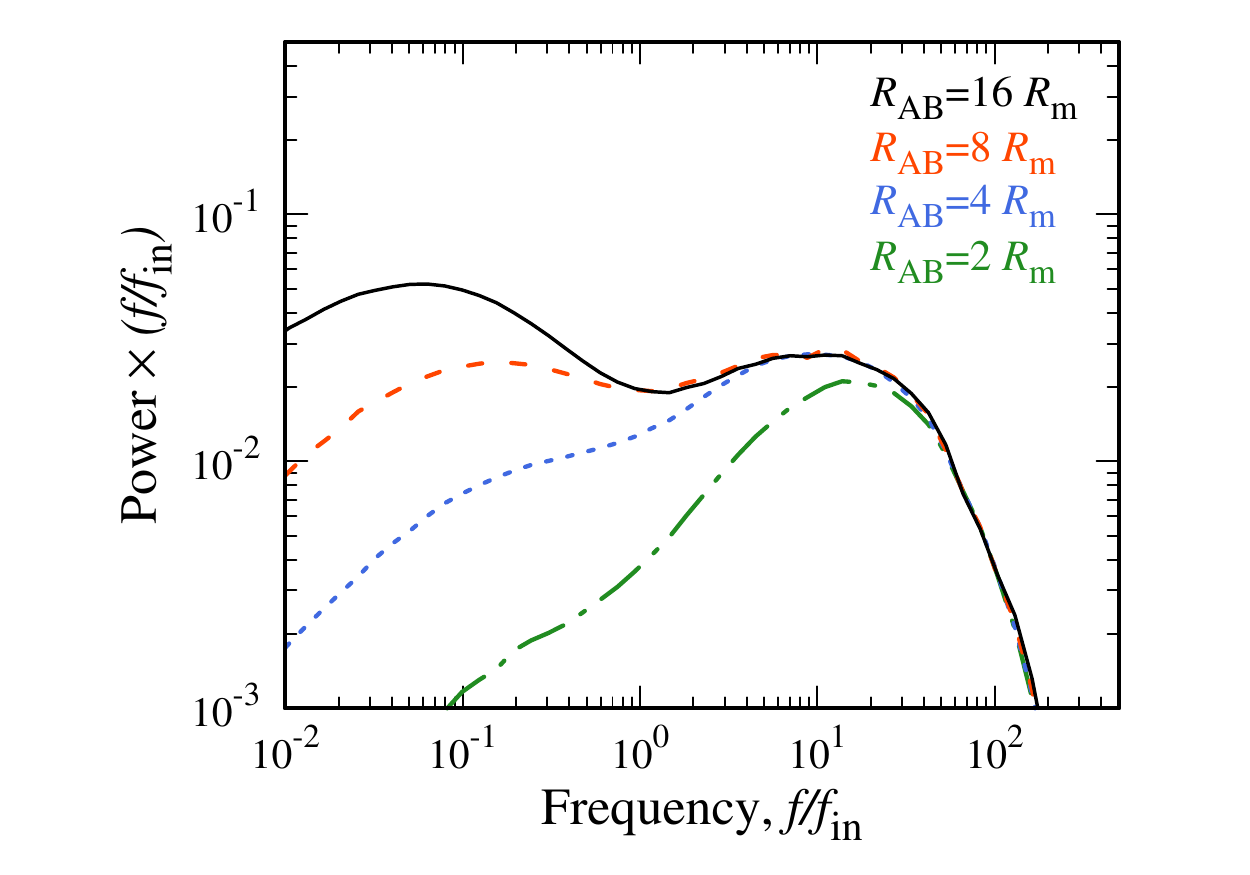}
\caption{ PDS modeled for a disk with a constant inner radius and a RPD zone with outer radius $R_\text{AB}$. 
Frequencies are normalized to the viscous frequency at the inner radius. 
We associate the increase of the RPD zone size with the increase of the luminosity as given by Eq.~\eqref{eq:r_ab}. 
For comparison with observed PDS, see Figure \ref{fig:pds}. 
 }
\label{fig:pds_model}
\end{figure}

Examples of the modeled PDS for a disk with a RPD inner region are shown in Figure \ref{fig:pds_model}. They were done for different sizes of RPD zone while the inner radius was kept constant. 
The modeling results reveal that the variability is dominated by the RPD inner disk.  
There are two components of power: a bump at low frequencies that depends strongly on the size of the RPD region and a bump at high frequencies with a roughly constant power. 
We see that the model PDSs provide good qualitative match to the observation ones (see Figure \ref{fig:pds}). 

Interestingly, the variability at low Fourier frequencies (below $\sim10$ Hz) in \gro\ is suppressed at high mass accretion rates and the suppression starts as soon as type-II bursts start to appear in the light curve of the source. 
Note that numerical simulations of accretion instability of RPD part of a disk \citep{Cannizzo1997} show an appearance of QPOs, which finally end up in a type-II burst.
Typical time scale of the oscillations is $t_{\rm osc}\sim 10-20$\,s.
We suggest that these oscillations suppress the fluctuations of the mass accretion rate at frequencies below $1/t_{\rm osc}$ originating from the outer parts of the accretion disk. 
As a result, the fluctuations observed at high mass accretion rates in \gro\ are produced predominantly in the inner RPD disk. 

At the highest luminosities, the changes in the shape of the low-frequency bump suggest that some additional processes at the outer edge of the RPD region shape the variability. The stability of the turn-over around $1$~Hz and the high-frequency QPOs remain without an explanation within the framework of the simplified model.

As noted above, the location and the shape of the high-frequency bump in the model PDS remain almost the same when the inner radius is fixed. The constant high-frequency edge confirms the assumption that there is no added noise above the maximal (Keplerian) frequency generated at the inner radius. 
Thus it is justified to relate the break frequency to the Keplerian frequency of the inner radius even in the RPD part of the disk. 

One complication to the observed variability may come from the hypothesized phenomenon of photon bubbles. They are thought to occur in both the accretion mound/column \citep[see, e.g.,][]{Pringle1972, Klein1996a} and in the RPD region of the disk \citep[e.g.][]{Gammie1998,Begelman2006}. However, there has not been clear evidence of their presence in PDS \citep{Revnivtsev2015}. Moreover, at least for the accretion column, the timescales are most likely outside the frequency range we examine, so we consider the scenario outlined above more plausible.

\subsection{Evolution of the inner disk edge}
\label{sec:depend}

As the luminosity increases, the inner radius derived from the Alfv\'en radius decreases as $R\subt{m}\propto L^{-2/7}$ (see Eq.~\eqref{eq:r_m}). If the break frequency is associated to the Keplerian frequency at the magnetospheric radius, we get the break frequency  luminosity-dependence $f\subt{b}\propto R\subt{m}^{-3/2}\propto L^\gamma$ with $\gamma=3/7\approx 0.43$. In \gro, however, the measured dependence of $\gamma=0.21\pm 0.02$, deviates from this classical relation. To the best of our knowledge, such a weak luminosity-dependence has not been observed before. In addition to that, the behavior between the outbursts may also be different; in the first outburst, the break frequency actually appears to be almost independent of the luminosity. 

The magnetospheric radius and its luminosity-dependence have been long discussed in the literature. A series of papers by \citet{Ghosh1977} and  \citet{Ghosh1979a,Ghosh1979,Ghosh1992} has become rather influential in the field of XRPs. In the context of an accretion disk truncated in the RPD zone, \citet{white1988} gave $\gamma=0.23$ based on a model by Ghosh \& Lamb. This theoretical dependence appears to be rather close to our measurement, but the choice of high magnetic diffusivity by Ghosh \& Lamb has been questioned by \citet{Wang1995}. Moreover, \cite{Bozzo2009} compared the models by Ghosh \& Lamb and Wang and discussed that the observations are not in good agreement with the magnetospheric radius behavior predicted in either of the theories. They concluded that although the Wang model is an improvement to the Ghosh \& Lamb model, it still has its own caveats due to which it cannot be straightforwardly applied. 

More recently, \citet{Chashkina2017,Chashkina2019} examined super-Eddington accretion onto magnetized NSs and found that in the RPD regime, the magnetospheric radius becomes almost independent of the mass accretion rate and larger than an inner radius expected for a GPD regime. In particular, \citet[][see their Fig. 14]{Chashkina2019} model gives a power-law dependence of $R\subt{m}$ on luminosity with index $\delta$ in the range between $-0.15$ and $-0.04$ for a dipole field and RPD inner disk, which would translate to $\gamma\approx 0.06-0.22$ consistent with the measured values. 
We note, however, that all aforementioned studies give $\xi$ an order of magnitude larger than that inferred from the PDS break frequencies assuming that they are Keplerian. 

As discussed in Sect.~\ref{sec:inner}, the magnetic field configuration affects the inner radius as well.
If the NS magnetic field is purely quadrupolar, the Keplerian frequency at the disk inner GPD  radius scales with luminosity with $\gamma=3/11\approx 0.27$, slightly deviating from what we observe. A small value of $\xi_{\rm q}$, on the other hand, may result from the specific field configuration having small values close to the orbital plane and allowing the disk to penetrate closer to the NS surface.

A weak dependence has also been found in the accretion disk simulations of \citet{Kulkarni2013} with $\gamma=0.3$. As a possible cause for such a dependence they mention the deformation of the magnetosphere by the accretion disk.

\section{Conclusions}
\label{sec:conclusions}

We examined the \xte\ observations of the Bursting Pulsar \gro\ and investigated how the power density spectrum (PDS) can be used as an independent method to examine the accretion disk properties of this source which has high luminosity and an intermediate magnetic field. We note that the observed PDS has a peculiar shape consisting of two bumps, thus deviating from a continuous power law from low to high frequencies predicted by perturbation propagation model for a gas-pressure dominated (GPD) accretion disk. The presence of strong variability at high frequencies indicates that the disk extends close to the NS surface. In this case, the disk should have a radiation-pressure dominated (RPD) inner region, which would affect the generation of variability in the disk. We modeled PDS using the perturbation propagation model for a disk with a RPD inner region and showed that during the major outbursts of \gro, the model PDSs are indeed in qualitative agreement with the observed ones. This suggests that the RPD region is present throughout the main outburst decay phase, but not in the mini-outbursts that follow. We note that a similar phenomenology is expected to be observed in ultraluminous X-ray sources hosting NSs as compact object and the hypothesis outlined above can be tested with high quality observations of these objects.

We measured the PDS break frequency as a function of luminosity during both outbursts and found a flatter dependence of the magnetospheric size on the luminosity than expected for a GPD disks.  
Associating the observed PDS high-frequency break with the Keplerian frequency of the inner radius of the disk truncated by the NS magnetic field, allowed us to estimate the inner disk radius which turned out to be much smaller than that expected for the dipole magnetic field of the strength determined from the cyclotron line.  Both small inner radius and its weak dependence on the luminosity can be explained by the magnetic field dominated by the quadrupole component and truncating the disk in the RPD region of the disk. Further research on both observational and theoretical sides is required to form a coherent picture of the system and how it relates to other X-ray pulsars.

\begin{acknowledgements}
This work was supported by the grant 14.W03.31.0021 of the Ministry of Science and Higher Education of the Russian Federation.  
VD thanks the Deutsches Zentrum for Luft- und Raumfahrt (DLR) and Deutsche Forschungsgemeinschaft (DFG) for financial support.  
VFS acknowledges the support from the DFG grant WE 1312/51-1 and from the travel grant of the German Academic Exchange Service (DAAD, project 57405000). 
This research was also supported by the Academy of Finland travel grants 317552 (JM, SST) and 322779 (JP), by the Netherlands Organization for Scientific Research Veni Fellowship (AAM) and by the V\"ais\"al\"a Foundation (SST).
The authors would like to acknowledge networking support by the COST Actions CA16104 and CA16214.
The authors are also grateful to Anna Chashkina, Pavel Abolmasov, Sergey Molkov and Phil Uttley for the valuable discussions.
\end{acknowledgements}

\bibliographystyle{aa}
\bibliography{library}

\end{document}